\documentclass[12pt,aps,prd,preprint,onecolumn]{revtex4}
\usepackage{amssymb}
\usepackage{amsmath}
\usepackage{graphics}
\usepackage{latexsym}
\usepackage{array}
\usepackage[activeacute,english]{babel}
\usepackage[latin1]{inputenc}
\usepackage[dvips]{graphicx}
\usepackage{color}

\newcommand{\bn}{\begin{eqnarray}}
\newcommand{\en}{\end{eqnarray}}
\newcommand{\be}{\begin{equation}}
\newcommand{\ee}{\end{equation}}

\newcommand{\bc}{\begin{center}}
\newcommand{\ec}{\end{center}}

\newcommand{\ket}[1]{\ensuremath{\left|#1\right\rangle}}
\newcommand{\bra}[1]{\ensuremath{\left\langle#1\right|}}

\newcommand{\abs}[1]{\left\vert#1\right\vert}
\newcommand{\crea}[2]{\hat{#1}^{\dag}_{#2}}
\newcommand{\des}[2]{\hat{#1}_{#2}}
\newcommand{\esp}[1]{\left[#1\right]}
\newcommand{\corc}[1]{\left(#1\right)}
\newcommand{\llav}[1]{{\left\{#1\right\}}}
\newcommand{\Dirac}{(i\gamma^{\mu}\partial_{\mu}-m)}

\begin{document}

\title{\Large \bf Majorana neutrino oscillations in vacuum}

\author{Y. F. P\'erez}
\email{yuber.perez@uptc.edu.co}

\affiliation{Escuela de F\'isica, Universidad Pedag\'ogica y
Tecnol\'ogica de Colombia, Tunja, Colombia}

\author{C. J. Quimbay\footnote{Associate researcher of Centro
Internacional de F\'{\i}sica, Bogot\'a D.C., Colombia.}}
\email{cjquimbayh@unal.edu.co}

\affiliation{Departamento de F\'{\i}sica, Universidad Nacional de Colombia.\\
Ciudad Universitaria, Bogot\'{a} D.C., Colombia.}

\date{\today}

\begin{abstract}
In the context of a type I seesaw scenario which leads to get
light left-handed and heavy right-handed Majorana neutrinos, we
obtain expressions for the transition probability densities
between two flavor neutrinos in the cases of left-handed and
right-handed neutrinos. We obtain these expressions in the context
of an approach developed in the canonical formalism of Quantum
Field Theory for neutrinos which are considered as superpositions
of mass-eigenstate plane waves with specific momenta. The
expressions obtained for the left-handed neutrino case after the
ultra-relativistic limit is taking lead to the standard
probability densities which describe light neutrino oscillations.
For the right-handed neutrino case, the expressions describing
heavy neutrino oscillations in the non-relativistic limit are
different respect to the ones of the standard neutrino
oscillations. However, the right-handed neutrino oscillations are
phenomenologically restricted as is shown when the propagation of
heavy neutrinos is considered as superpositions
of mass-eigenstate wave packets.\\

\vspace{0.4cm} \noindent {\it{Keywords:}} {Majorana fermions,
Majorana neutrino oscillations, transition probability,
non-relativistic and ultra-relativistic approximations.}

\vspace{0.4cm} \noindent {\it{PACS Codes:}}  11.10.-z, 13.15.+g,
14.60.-z.

\end{abstract}

\maketitle


\section{Introduction}\label{sec:01}

Neutrino physics is a very active area of research which involves
some of the most intriguing problems in particle physics. The
nature of neutrinos and the origin of the small mass of neutrinos
are two examples of these kind of problems. Since neutrinos are
electrically neutral, the nature of these elementary particles can
be Majorana or Dirac fermions. The first possibility, i. e.
neutrinos being Majorana fermions was introduced by Etore Majorana
\cite{M37} when he suggested that massive neutral fermions with
specific momenta have associated only two helicity states implying
that neutrinos and anti-neutrinos are the same particles. The
second possibility implies that Dirac neutrinos are described by
four-component spinorial fields which are different from spinorial
fields describing anti-neutrinos. In this work, we will consider
neutrinos as Majorana fermions which is favored by simplicity
because they have only two degrees of freedom \cite{RM02,GK07}.

Direct and indirect experimental evidences show that neutrinos are
massive fermions with masses smaller than $1$ eV \cite{PDG}. The
most accepted way to generate neutrino masses is by mean of the
seesaw mechanism \cite{Bot}. Mass for neutrinos is a necessary
ingredient to understand the oscillations between neutrino flavor
states which have been observed experimentally \cite{PDG}.
Neutrino oscillations are originated by the interference between
mass states whose mixing generates flavor states. This phenomenon
means that a neutrino created in a weak interaction process with a
specific flavor can be detected with a different flavor. Neutrino
oscillations were first described by Pontecorvo \cite{Ponteco} as
an extension for the leptonic sector of the strange oscillations
observed in the neutral Kaon system. Neutrino oscillations can be
described in context of Quantum Mechanics
\cite{Ponteco1}-\cite{Zralec} as an application of the two level
system \cite{Sassa0}.

Description of neutrino oscillations in the context of Quantum
Field Theory (QFT) is a very well studied topic
\cite{Blasone}-\cite{Aknme}. In the literature it is possible to
find two kinds of QFT models describing neutrino oscillations:
intermediate models and external models \cite{Beuthe}. In the
framework of intermediate models Sassaroli developed a model based
in an interacting Lagrangian density which includes the coupling
between two flavor fields \cite{Sassa1,Sassa2}. This model was
framed by Beuthe as a hybrid model owing to it goes half-a-way to
QFT \cite{Beuthe}. Sassaroli model was first developed for a
coupled system of two Dirac equations \cite{Sassa1} and then it
was extended for a coupled system of two Majorana ones
\cite{Sassa2}. The probability amplitude of transition between two
neutrino flavor states for these two systems \cite{Sassa1,Sassa2}
was obtained starting from flavor states which are used on the
standard treatment of neutrino oscillations.

The standard definition of flavor states can originate some
possible limitations in the description of neutrino oscillations
as was observed by Giunti et al. \cite{Giunti2}. Specifically, in
reference \cite{Giunti2} it was shown that flavor states can
define an approximate Fock space of weak states in the following
two cases: (i) In the extremely relativistic limit, i. e. if
neutrino momentum is much larger than the maximum mass eigenvalue
of a neutrino mass state; (ii) for almost degenerated neutrino
mass eigenvalues, i. e. if the differences between neutrino mass
eigenvalues are much smaller than the neutrino momentum. The first
case leads to the standard definition of flavor states. The second
case has associated a real mixing matrix which is restricted to a
specific interaction process. Additionally these authors have
proposed that oscillations can be described appropriately for
ultra-relativistic and non-relativistic neutrinos by defining
appropriate flavor states which are superpositions of mass states
weighted by their transitions amplitudes in the process under
consideration \cite{Giunti2}.

By considering the limitations mentioned in the last paragraph it
was set down by Beuthe in \cite{Beuthe} that Sassaroli hybrid
model can only be applied consistently if lepton flavor wave
functions are considered as observable and the ultra-relativistic
limit is taken into account. On the other hand, the Sassaroli
model describing Majorana neutrino oscillations \cite{Sassa1} was
developed without considering the four-momentum conservation for
neutrinos which implies the existence of a specific momentum for
every neutrino mass state.

The main goal of this work is to study neutrino oscillations in
vacuum between two flavor states considering neutrinos as Majorana
fermions and to obtain the probability densities of transition for
left-handed neutrinos (ultra-relativistic limit) and for
right-handed neutrinos (non-relativistic limit). This work is
developed in the context of a type I seesaw scenario which leads
to get light left-handed and heavy right-handed Majorana
neutrinos. In this context, we perform an extension of the model
developed by Sassaroli in which the Majorana neutrino oscillations
are obtained for the case of flavor states constructed as
superpositions of mass states \cite{Sassa1}. Our extension
consists in considering neutrino mass states as plane waves with
specific momenta. The model that we consider in this work, which
is developed in the canonical formalism of Quantum Field Theory,
has the advantage that in the same theoretical treatment it is
possible to study neutrino oscillations for light neutrinos and
for heavy neutrinos. To do this, we first perform the canonical
quantization procedure for Majorana neutrino fields of definite
masses and then we write the neutrino flavor states as
superpositions of mass states using quantum field operators. Next
we calculate the probability amplitude of transition between two
different neutrino flavor states for the light and heavy neutrino
cases and we establish normalization and boundary conditions for
the probability density. These probability densities for the
left-handed neutrino case after the ultra-relativistic limit is
taking lead to the standard probability densities which describe
light neutrino oscillations. For the right-handed neutrino case,
the expressions describing heavy neutrino oscillations in the
non-relativistic limit are different respect to the ones of the
standard neutrino oscillations. However, the right-handed neutrino
oscillations are phenomenologically restricted as is shown when
the propagation of heavy neutrinos is considered as superpositions
of mass-eigenstate wave packets \cite{Kim}. The oscillations do
not take place in this case because the coherence is not
preserved: in other words, the oscillation length is comparable or
larger than the coherence length of the neutrino system
\cite{Kim}.

The content of this work has been organized as follows: In section
2, after establishing the Majorana condition, we obtain and solve
the two-component Majorana equation for a free fermion; in section
3, we consider a type I seesaw scenario which leads to get light
left-handed neutrinos and heavy right-handed neutrinos; in section
4, we obtain the Majorana neutrino fields with definite masses,
then we carry out the canonical quantization procedure of these
Majorana neutrino fields and we obtain relation between neutrino
flavor states and neutrino mass states using operator fields; in
section 5, we determine the probability density of transition
between two left-handed neutrino flavor states, additionally we
establish normalization and boundary conditions and then we obtain
left-handed neutrino oscillations for ultra-relativistic light
neutrinos; in section 6, we study the right-handed neutrino
oscillations for non-relativistic heavy neutrinos; finally, in
section 7 we present some conclusions.


\section{Two-component Majorana equation} \label{Majeq}

In 1937 Ettore Majorana proposed a symmetric theory for electron
and positron through a generalization of a variational principle
for fields which obey Fermi-Dirac statistics \cite{M37}. When this
theory is applied to a neutral fermion which has a specific
momentum then there exist only two helicity states. The Majorana
theory implies that it does not exist antiparticles associated to
these fermions, i. e. Majorana fermions are their own
antiparticles. For convenience we study the equation of motion for
neutral fermions but using the two-component theory developed by
Case in \cite{Case}.

In contrast with a Dirac fermion, a Majorana fermion can only be
described by a two-component spinor. To show it we consider a free
relativistic fermionic field $\psi$ of mass $m$ described by the
Dirac equation $\Dirac\psi=0$, where Dirac matrixes $\gamma^\mu$
obey the anticonmutation relations
$\llav{\gamma^\mu,\gamma^\nu}=2g^{\mu\nu}$ and metric tensor
satisfies $g_{\mu\nu}=g^{\mu\nu}=\text{diag}(1,-1,-1,-1)$. Using the
chirality matrix given by $\gamma^5\equiv
i\gamma^0\gamma^1\gamma^2\gamma^3$, the left- and right-handed
chiral projections of the fermion field $\psi$ are
$\psi_{R,L}=\frac{1}{2}\corc{1\pm\gamma^5}\psi$, respectively. If we
write the Dirac matrixes projected on the chiral subspace as
$\gamma^\mu_{\pm}=\frac{1}{2}\corc{1\pm\gamma^5}\gamma^\mu$, we
obtain that the coupled equations for the chiral components of the
fermionic field $\psi$ are given by
\begin{subequations}\label{eq:EQ}
\begin{align}
i\gamma^\mu_+\partial_\mu\psi_L=m\psi_R,\label{eq:EQ1}\\
i\gamma^\mu_-\partial_\mu\psi_R=m\psi_L.\label{eq:EQ2}
\end{align}
\end{subequations}

We introduce the charge conjugation operation that will allow us to
describe Majorana fermions. The charged conjugated field (or
conjugated field) $\psi^c$ is defined as
\begin{equation}
\psi^c=\mathcal{\hat C}\bar\psi^T,
\end{equation}
where the charge conjugation operator $\mathcal{\hat C}$ satisfies
the properties $\mathcal{\hat C}(\gamma^\mu)^T\mathcal{\hat
C}^{-1}=-\gamma^\mu$, $\mathcal{\hat C}^{-1}=\mathcal{\hat C}^\dag$,
$\mathcal{\hat C}^T=-\mathcal{\hat C}$  \cite{GK07}. Using these
properties we find that the conjugated field $\psi^c$ obeys the
Dirac equation $\Dirac\psi^c=0$. As $\psi$ describes a fermion with
a specific charge, its conjugated field $\psi^c$ represents a
fermion with an opposite charge and with the same mass, i. e.
$\psi^c$ describes the antifermion of $\psi$. The Dirac equation for
$\psi^c$ should be projected on the chiral subspace and for this
reason it is necessary to remember that $\mathcal{\hat
C}(\gamma^5)^T=\gamma^5\mathcal{\hat C}$ \cite{GK07}. So the coupled
equations for the chiral components of the conjugated field $\psi^c$
are
\begin{subequations}\label{eq:ECQ}
\begin{align}
i\gamma^\mu_+\partial_\mu(\psi_R)^c=m(\psi_L)^c,\label{eq:ECQ1}\\
i\gamma^\mu_-\partial_\mu(\psi_L)^c=m(\psi_R)^c.\label{eq:ECQ2}
\end{align}
\end{subequations}
We observe that the chiral components of the fermionic field $\psi$
under charge conjugation $(\psi^c)_L$, $(\psi^c)_R$ and the chiral
components of the conjugated field $(\psi_R)^c$, $(\psi_L)^c$ are
related by $(\psi_L)^c=(\psi^c)_R$, $(\psi_R)^c=(\psi^c)_L$, showing
how the charge-conjugation operation changes the chirality of
fields.

We define the Majorana condition by taking the fermionic field as
proportional to the conjugated field
\begin{align}\label{eq:CM}
\psi \equiv \xi\psi^c,
\end{align}
where the proportional constant is a complex phase factor of the
form $\xi\equiv e^{i\alpha}$ which plays an important role on
applications of Majorana theory. The equality (\ref{eq:CM}) implies
that Majorana fermions are their own antiparticles. Now the chiral
components of the Majorana field satisfy
\begin{align}\label{eq:CMQ}
\psi_L=\xi\ \mathcal{\hat C}\bar \psi_R^T,\quad \psi_R=\xi\
\mathcal{\hat C}\bar \psi_L^T.
\end{align}
So we can write equations (\ref{eq:EQ1}) and (\ref{eq:EQ2}) in the
form
\begin{subequations}\label{eq:EM}
\begin{align}
i\gamma^\mu_+\partial_\mu\psi_L&=m\xi\mathcal{\hat C}\bar \psi_L^T,\label{eq:EM1}\\
i\gamma^\mu_-\partial_\mu\mathcal{\hat C}\bar
\psi_L^T&=m\xi^*\psi_L.\label{eq:EM2}
\end{align}
\end{subequations}
If we apply the Majorana condition (\ref{eq:CMQ}) into the
equations (\ref{eq:ECQ1}) and (\ref{eq:ECQ2}), we obtain equations
(\ref{eq:EM1}) and (\ref{eq:EM2}). Additionally we can observe
that equations (\ref{eq:EM1}) and (\ref{eq:EM2}) are related to
themselves by means of a complex conjugation. In this way, we have
gone from four coupled equations describing a fermion and its
antifermion to two decoupled equations describing a left-handed
chiral field $\psi_L$ and a right-handed chiral field $\psi_R$.
Due to the fact that the right-handed chiral field can be
constructed from the left-handed chiral field \cite{Case}, as it
is shown in (\ref{eq:CMQ}), now we have only an independent field
given by $\psi_L$. For the last fact, we will be able to describe
Majorana fermion by means of field $\psi_L$ which now has two
components. To verify this sentence we rewrite equation
(\ref{eq:EM1}) as
\begin{align}\label{eq:EQM1}
i\mathcal{\hat
C}\gamma^0\gamma^\mu_+\partial_\mu\psi_L=m\xi\psi_L^*.
\end{align}
If we define
\begin{equation}\label{eq:MM}
\eta^\mu \equiv\mathcal{\hat C}\gamma^0\gamma^\mu_+,
\end{equation}
and if we take $\phi\equiv\psi_L$, then equation (\ref{eq:EQM1}) can
be written as
\begin{align}\label{eq:EMDC1}
i\eta^\mu\partial_\mu\phi=m\xi\phi^*,
\end{align}
which is known as the Majorana equation. This equation in which a
particle is indistinguishable from its antiparticle has two
components because the matrixes $\gamma^\mu_\pm$ are projected on
the chiral subspaces of two components. The matrixes $\eta^\mu$ are
called Majorana matrixes and these should not be confused with the
Dirac matrixes written in Majorana representation.

Now we are interested in knowing the kind of relations that
Majorana matrixes $\eta^\mu$ obey. So we first apply definition
(\ref{eq:MM}) into equation (\ref{eq:EM2}) and we obtain
$-i\eta^{\mu*}\partial_\mu\phi^*=m\xi^*\phi$, with
$\eta^{\mu*}=\gamma^\mu_-\gamma^0\mathcal{\hat C}$. Then we apply
$-i\eta^{\nu*}\partial_\nu$ into (\ref{eq:EMDC1}) and we have
$\eta^{\nu*}\eta^\mu\partial_\nu\partial_\mu\phi-m^2\phi=0 $ or
its equivalent
$\frac{1}{2}\llav{\eta^{\mu*}\eta^\nu+\eta^{\nu*}\eta^\mu}\partial_
\nu\partial_\mu\phi-m^2\phi=0$, where we have used
$\abs{\xi}^2=1$. Accordingly, Majorana matrixes should satisfy
relations $\eta^{\mu*}\eta^\nu+\eta^{\nu*}\eta^\mu=-2g^{\mu\nu}$
and then the field $\phi$ is satisfying the Klein-Gordon equation
given by $\corc{\partial^\mu\partial_\mu+m^2}\phi=0$.

In this work we have taken a particular representation of matrixes
$\eta^\mu$ which has permitted us to write the two-component
Majorana equation in the form given by (\ref{eq:EMDC1}). Now we
can consider a matrix $A$ which satisfies the following relations
\cite{Case}
\begin{align}\label{eq:PA}
A\sigma^{i*}A^{-1}=-\sigma^i,\quad A=A^\dag=A^{-1}=-A^T,
\end{align}
where $\sigma^i$ represents Pauli matrixes in a given
representation. We take $\eta^\mu=i\sigma_2\bar{\sigma}^\mu$,
where $\bar{\sigma}^\mu\equiv(\mathbb{I},-\vec{\mathbf{\sigma}})$
being $\mathbb{I}$ the unit matrix $2\times 2$ and
$\vec{\mathbf{\sigma}}=(\sigma_1,\sigma_2,\sigma_3)$ Pauli
matrixes. Since $\sigma_2$ satisfies  properties (\ref{eq:PA}), we
have taken $A\equiv\sigma_2$. So the equation (\ref{eq:EMDC1}) can
be written as
\begin{align}\label{eq:EMDC}
i\bar{\sigma}^\mu\partial_\mu\phi+im\xi\sigma_2\phi^*=0.
\end{align}
This equation is the well known two-component Majorana equation
\cite{Pal,Marsch}, which will be solved in next subsection.


\subsection{Canonical quantization for Majorana field}

With the purpose of studying the canonical quantization for the
Majorana field we will obtain the free-particle solution of
equation (\ref{eq:EMDC}). On the outset, we consider bi-spinors
$\chi$ which obey the following relations
\begin{subequations}
\begin{align}
\frac{\vec{\mathbf{\sigma}}\cdot\vec{p}}{\abs{\vec{p}}}
\chi^h(\vec{p})&=h\chi^h(\vec{p}),\\
-i\sigma_2(\chi^h(\vec{p}))^*&=h\chi^{-h}(\vec{p}),
\end{align}
\end{subequations}
where these bi-spinors correspond to helicity eigenstates. If we
take the momentum in spherical coordinates
$\vec{p}=\abs{\vec{p}}(\sin\theta\cos\varphi, \sin\theta\sin\varphi,
\cos\theta)$, then the helicity operator has the form
\begin{align}
\frac{\vec{\mathbf{\sigma}}\cdot\vec{p}}{\abs{\vec{p}}}=\begin{pmatrix}
                                 \cos\theta & \sin\theta e^{-i\varphi}\\
                                 \sin\theta e^{i\varphi} & -\cos\theta
                        \end{pmatrix}.
\end{align}
We choose the following representation for these bi-spinors
\begin{align}
\chi^+(\vec{p})=\begin{pmatrix}
            \cos\frac{\theta}{2}\\
            \sin\frac{\theta}{2}e^{i\varphi}
                \end{pmatrix},\quad
\chi^-(\vec{p})=\begin{pmatrix}
          -\sin\frac{\theta}{2}e^{-i\varphi}\\
           \cos\frac{\theta}{2}
                 \end{pmatrix}.
\end{align}
We can prove that the following solution satisfies the two-component
Majorana equation (\ref{eq:EMDC})
\begin{align} \label{eq:SeM}
\phi^h_{\vec{p}}(x)=\sqrt{\frac{E-h\abs{\vec{p}}}{2E}}\chi^h(\vec{p})
e^{-ipx}-h\sqrt{\frac{E+h\abs{\vec{p}}}{2E}}\chi^{-h}(\vec{p})e^{ipx},
\end{align}
with $px\equiv p_\mu x^\mu=Et-\vec{p}\cdot\vec{x}$. We observe that
Majorana field can be written as superposition of positive and
negative energy states.

The Lagrangian density which describes a free two-component
Majorana field is given by
\begin{align} \label{eq:Mld}
\mathcal{L}_M=\phi^{\dag}i\bar\sigma^{\mu}\partial_{\mu}\phi-
\frac{m\xi}{2}\corc{\phi^{T}i\sigma_2\phi-\phi^{\dag}i\sigma_2\phi^*},
\end{align}
where the two-component Majorana field $\phi$ and its conjugated
field $\phi^{\dag}$ behave as Grassmann variables. It is very easy
to prove that the two-component Majorana equation (\ref{eq:EMDC})
can be obtained from the Lagrangian density (\ref{eq:Mld}) using
the Euler-Lagrange equation. Additionally, we can obtain the
following energy-momentum tensor from (\ref{eq:Mld}), $
T_{\mu\nu}=\phi^{\dag}i\bar\sigma_{\mu}\partial_{\nu}\phi\notag
-g_{\mu\nu}\esp{\phi^{\dag}i\bar\sigma^{\lambda}\partial_{\lambda}
\phi-\frac{m}{2}\corc{\phi^{T}i\sigma_2\phi-\phi^{\dag}i\sigma_2\phi^*}}
$. Following the standard canonical quantization procedure, we now
consider the Majorana field $\phi$ and its conjugated field
$\phi^{\dag}$ as operators which satisfy the usual canonical
anticonmutation relations given by $
\llav{\hat\phi_{\alpha}(\vec{r},t),\hat\phi_{\beta}(\vec{r}\,',t)}=
\llav{\hat\phi^{\dag}_{\alpha}(\vec{r},t),\hat\phi^{\dag}_{\beta}(\vec{r}\,',t)}=0$,
$
\llav{\hat\phi_{\alpha}(\vec{r},t),\hat\phi^{\dag}_{\beta}(\vec{r}\,',t)}=
\delta_{\alpha\beta}\delta^3(\vec{r}-\vec{r}\,')$, where
$\alpha,\beta=1,2$. Using the Heisenberg equation for the Majorana
field $
i\partial_t\hat\phi_\alpha(\vec{r},t)=\esp{\hat\phi_\alpha(\vec{r},t),\hat{
H}}$, we can obtain its corresponding Majorana equation
(\ref{eq:EMDC}). By means of the energy-momentum tensor it is
possible to prove that the Hamiltonian operator can be written as
$\hat{H}=\int\
d^3x\corc{\hat{\phi}^{\dag}i\vec{\mathbf{\sigma}}\cdot
\nabla\hat{\phi}+\frac{m\xi}{2}\corc{\hat{\phi}^{T}i\sigma_2\hat{\phi}-
\hat{\phi}^{\dag}i\sigma_2\hat{\phi}^*}}$. The expansion in a
Fourier series for the Majorana field operator is
\cite{Giunti2}-\cite{Case}
\begin{align}\label{eq:EONM}
\hat\phi(x)=\int\frac{d^3p}{(2\pi)^{3/2}(2E)^{1/2}}&\sum_{h=\pm1}
\left[\sqrt{E-h\abs{\vec{p}}}\,\des{a}{}(\vec{p},h)\chi^{h}(\vec{p})
e^{-ip\cdot x}\right.\notag\\ &\qquad\quad\left.-h\sqrt{E+h\abs{\vec{p}}}\,
\crea{a}{}(\vec{p},h)\chi^{-h}(\vec{p})e^{ip\cdot x}\right],
\end{align}
where we have used the free-particle solution (\ref{eq:SeM}) and
operators $\des{a}{}(\vec{p},h)$, $\crea{a}{}(\vec{p},h)$ which
satisfy the anticonmutation relations
$\llav{\des{a}{}(\vec{p},h),\des{a}{}(\vec{p}\,',h')}=
\llav{\crea{a}{}(\vec{p},h),\crea{a}{}(\vec{p}\,',h')}=0$,
$\llav{\des{a}{}(\vec{p},h),\crea{a}{}(\vec{p}\,',h')}=
\delta_{h,h'}\delta^3(\vec{p}-\vec{p}\,')$. Then we can identify
$\des{a}{}(\vec{p},h)$ as the annihilation operator and
$\crea{a}{}(\vec{p},h)$ as the creation operator of a Majorana
fermion with momentum $\vec{p}$ and helicity $h$.


\section{Masses for Majorana neutrino fields}

The most accepted way to generate neutrino masses is through the
seesaw mechanism. In this section we consider a type I seesaw
scenario which leads to get light left-handed neutrinos and heavy
right-handed neutrinos. For the case of two neutrino generations,
a Dirac-Majorana mass term is given by \cite{Bilpet}
\begin{equation}\label{eq:DMT}
\mathcal{L}^{M+D}_Y=\frac{1}{2} \bar N_L \mathcal{\hat C}
M^{M+D}_\nu N_L + H.c.,
\end{equation}
where $H.c.$ represents the hermitic conjugate term, $N_L$ is the
vector of flavor neutrino fields written as
\begin{align}\label{eq:NL1}
N_L=\begin{pmatrix}
       \nu_L\\
    \mathcal{\hat C} \bar \nu^T_R
      \end{pmatrix}
   =\begin{pmatrix}
       \nu_L\\
    \nu^c_R
      \end{pmatrix},
\end{align}
where $\nu_L$ represents a doublet of left-handed neutrino fields
active under the weak interaction and $\nu^c_R$ represents a
doublet of right-handed Majorana neutrino fields non active
(sterile) under the weak interaction. These doublets are given by
\begin{align}\label{eq:NL2}
\nu_L=\begin{pmatrix}
       \nu_{e_L}\\
       \nu_{\mu_L}
      \end{pmatrix};
   \,\,\,\nu^c_R=\begin{pmatrix}
       \nu^c_{e_R}\\
       \nu^c_{\mu_R}
      \end{pmatrix}.
\end{align}
In the Dirac-Majorana term (\ref{eq:DMT}), $M^{M+D}_\nu$ is a
$4\times4$ non-diagonal matrix of the form
\begin{align}\label{eq:MMMnu}
M^{M+D}_\nu=\begin{pmatrix}
   M^{\prime}_L & (M_D)^T\\
   M_D & M^{\prime}_R
  \end{pmatrix},
\end{align}
where $M_L$, $M_R$ and $M_D$ are $2\times2$ matrixes. The vector
of neutrino fields with definite masses $n_L$ can be written by
mean of a unitary matrix $U^\nu_L$ as follows
\begin{equation}\label{eq:FTUM}
N_L=U^\nu_L n_L,
\end{equation}
where $n_L$ has the form
\begin{align}\label{eq:NL3}
n_L=\begin{pmatrix}
       n_{1}\\
       n_{2}\\
      \end{pmatrix}=\begin{pmatrix}
       \nu_{1}\\
       \nu_{2}\\
       \nu_{3}\\
       \nu_{4}
      \end{pmatrix}.
\end{align}
The unitary matrix $U^\nu_L$ is chosen in such a way that the
non-diagonal matrix $M^{M+D}_\nu$ can be diagonalized through the
similarity transformation
\begin{equation}\label{eq:MMdiag}
(U^\nu_L)^{-1} M^{M+D}_\nu U^\nu_L = M_\nu,
\end{equation}
where $M_\nu$ is a diagonal matrix which is defined by
$(M_\nu)_{ab} = m_a \delta_{ab}$, where $a,b=1,2,3,4$. The masses
of the neutrino fields of definite masses $\nu_{a}$ are $m_a$,
with $a=1,2,3,4$.

The seesaw scenario is established imposing the following
conditions into the matrix (\ref{eq:MMMnu}): $M^{\prime}_L = 0$,
$(M_D)_{kj} \ll (M^{\prime}_R)_{kj}$, thus the matrix
$M^{M+D}_\nu$ is diagonalized as
\begin{equation}\label{eq:MMdiag}
(U^\nu_L)^{-1} M^{M+D}_\nu U^\nu_L = \begin{pmatrix}
   M_{l}& 0\\
   0 & M_{h}
  \end{pmatrix},
\end{equation}
where $M_{l}$ is the light neutrino mass matrix and $M_{h}$ is the
heavy neutrino mass matrix. If the unitary matrix $U^\nu_L$ is
expanding considering terms until of the order $(M^\prime_R)^{-1}
M_D$, the light and heavy neutrino mass matrixes can be written as
\begin{equation}\label{eq:MMlh}
M_{l}=\begin{pmatrix}
   m_1 & 0\\
   0 & m_2
  \end{pmatrix};
\,\,\,M_{h}=\begin{pmatrix}
   m_3 & 0\\
   0 & m_4
  \end{pmatrix}.
\end{equation}
The Dirac-Majorana mass term (\ref{eq:DMT}) can be written in
terms of the neutrino fields of definite masses $\nu_{a}$ as
\begin{equation}\label{eq:DMT1}
\mathcal{L}^{M+D}_Y=\frac{1}{2} \bar n_{1} \mathcal{\hat C} M_{l}
n_{1}+\frac{1}{2} \bar n_{2} \mathcal{\hat C} M_{h} n_{2}+ H.c.,
\end{equation}
where the matrixes $M_l$ and $M_h$ are given by (\ref{eq:MMlh})
and the doublets $n_{1_L}$ and $n_{2_L}$ are written as
\begin{align}\label{eq:NL4}
n_{1}=\begin{pmatrix}
       \nu_{1}\\
       \nu_{2}
      \end{pmatrix};
   \,\,\,n_{2}=\begin{pmatrix}
       \nu_{3}\\
       \nu_{4}
      \end{pmatrix}.
\end{align}
The neutrino fields of definite masses $\nu_{1}$ and $\nu_{2}$
have associate respectively the light masses $m_1 \sim
m_e^2/(f_{11} v_R)$ and $m_2 \sim m_\mu^2/(f_{22} v_R)$ and the
neutrino fields of definite masses $\nu_{3}$ and $\nu_{4}$ have
associate respectively the heavy masses $m_3 = f_{33} v_R$ and
$m_4 = f_{44} v_R$, where $v_R \rightarrow \infty$, $f_{ab}$ are
Yukawa couplings, $m_e$ is the electron mass and $m_\mu$ is the
muon mass.

As it will be shown in the next section, starting from the
Dirac-Majorana mass term
\begin{equation}\label{eq:DMT2}
\mathcal{L}^{M+D}_Y=\frac{1}{2} \bar \nu_L \mathcal{\hat C} M_{L}
\nu_L+\frac{1}{2} \bar \nu^c_R \mathcal{\hat C} M_{R} \nu^c_R+
H.c.,
\end{equation}
where $\nu_L$ and $\nu^c_R$ are the flavor doublets of
non-definite masses given by (\ref{eq:NL2}), while $M_L$ and $M_R$
are $2\times2$ non-diagonal matrixes, it will be possible to
obtain the Dirac-Majorana mass term (\ref{eq:DMT1}) after the
diagonalization of the matrixes $M_L$ and $M_R$.


\section{Mass and flavor neutrino states}

In the next we suppose that the Majorana fields  $\nu_{e_L}$ and
$\nu_{\mu_L}$ describe the active light left-handed neutrinos that
are produced and detected in the laboratory, while the Majorana
fields $\nu^c_{e_R}$ and $\nu^c_{\mu_R}$ describe the sterile
heavy right-handed neutrinos which there exist in a type I seesaw
scenario.

In the section (\ref{Majeq}) we have presented a lagrangian
density (\ref{eq:Mld}) which describes a free Majorana fermion.
This lagrangian density can be extended to describe a system of
two flavor left-handed neutrinos and two flavor right-handed
neutrinos with non-definite masses. Using the Dirac-Majorana mass
term given by (\ref{eq:DMT2}), the lagrangian density describing
this system is given by
\begin{align}\label{eq:MAC}
\mathcal{L}=\bar \nu_L i\bar\sigma^{\mu}\partial_{\mu}\nu_L+ \bar
\nu^c_R i\bar\sigma^{\mu}\partial_{\mu}\nu^c_R-\frac{1}{2} \bar
\nu_L M_L i\sigma_2 \nu_L -\frac{1}{2} \bar \nu^c_R M_R i\sigma_2
\nu^c_R + H.c.
\end{align}
where the non-diagonal mass matrixes $M_L$ and $M_R$ are written
as
\begin{align}\label{eq:MMML}
M_L=\begin{pmatrix}
   m_{\nu_{e_L}}e^{i\delta_1} & m_{\nu_{e_L}\nu_{\mu_L}}e^{i(\delta_1+\delta_2)}\\
   m_{\nu_{e_L}\nu_{\mu_L}}e^{i(\delta_1+\delta_2)} & m_{\nu_{\mu_L}} e^{i\delta_2}
  \end{pmatrix},
\end{align}
\begin{align}\label{eq:MMMR}
M_R=\begin{pmatrix}
   m_{\nu_{e_R}}e^{i\delta_3} & m_{\nu_{e_R}\nu_{\mu_R}}e^{i(\delta_3+\delta_4)}\\
   m_{\nu_{e_R}\nu_{\mu_R}}e^{i(\delta_3+\delta_4)} & m_{\nu_{\mu_R}} e^{i\delta_4}
  \end{pmatrix},
\end{align}
We observe that the form of the matrixes $M_L$ and $M_R$ is the
same. In the next, we will restrict to the left-handed Majorana
neutrinos, but the results are directly extended to the
right-handed Majorana neutrinos. From the Euler-Lagrange equations
we obtain that the coupled equation of motion for the flavor
left-handed neutrino fields $\nu_{e_L}$ and $\nu_{\mu_L}$ are
\begin{subequations}\label{eq:MAC2}
\begin{align}
i\bar\sigma^{\lambda}\partial_{\lambda}
\nu_{e_L}&=-im_{\nu_{e_L}}\sigma_{2}e^{i\delta_1}\nu_{e_L}^{*}-im_{\nu_{e_L}\nu_{\mu_L}}
\sigma_{2}e^{i(\delta_1+\delta_2)}\nu_{\mu_L}^{*},\\
i\bar\sigma^{\lambda}\partial_{\lambda}\nu_{\mu_L}&=-im_{\mu_L}
\sigma_{2}e^{i\delta_2}\nu_{\mu}^{*}-im_{\nu_{e_L}\nu_{\mu_L}}\sigma_{2}e^{i(\delta_1+\delta_2)}
\nu_{e_L}^{*},
\end{align}
\end{subequations}
respectively. We observe that flavor neutrino fields are coupled
by means of the parameter $m_{\nu_{e_L}\nu_{\mu_L}}$. With the
purpose of decoupling the equations of motion for the flavor
left-handed neutrino fields, now we consider the most general
unitary matrix $U_L$ given by
\begin{align}
 U_L=\frac{1}{\sqrt{1+\Lambda_L^2}}\begin{pmatrix}
                                \Lambda_L e^{-i\frac{\delta_1}{2}} &
                                e^{-\frac{i}{2}(\delta_1+\alpha_L)}\\
                -e^{-i\frac{\delta_2}{2}} & \Lambda_L e^{-\frac{i}{2}
                (\delta_2+\alpha_L)}
                               \end{pmatrix},
\end{align}
where the phases $e^{-i\delta_1}$ and $e^{-i\delta_2}$ appear as a
consequence of the Majorana condition (\ref{eq:CMQ}). The
definite-mass neutrino field doublet $n_1$ given by (\ref{eq:NL4})
is related to the flavor left-handed neutrino doublet $\nu_L$
given by (\ref{eq:NL2}) by mean of
\begin{equation}\label{eq:CMN}
\nu_L=U_L n_1.
\end{equation}
Without a lost of generality, we can change the phases of the
flavor left-handed neutrino fields by means of
$\nu_{e_L}\rightarrow \exp\{-\frac{i\delta_1}{2}\}\nu_{e_L}$ and
$\nu_{\mu_L}\rightarrow \exp\{-\frac{i\delta_2}{2}\}\nu_{\mu_L}$.
Thus there is just a phase $\exp\{-\frac{i\alpha_L}{2}\}$ that can
not be eliminated. So the matrix $U$ can be rewritten as
\begin{align}
 U_L=\frac{1}{\sqrt{1+\Lambda_L^2}}\begin{pmatrix}
                                \Lambda_L & e^{-\frac{i\alpha_L}{2}}\\
                -1 & \Lambda_L e^{-\frac{i\alpha_L}{2}}
                               \end{pmatrix}.
\end{align}
Now the diagonalization of the mass matrix (\ref{eq:MMML}) given
by
\begin{align}\label{eq:DONM}
 M_{D_L}=U_L^\dag M_L U_L=\text{diag}(m_1,m_2),
\end{align}
is valid for
\begin{equation} \label{eq:Lambda}
 \Lambda_L =\frac{m_{\nu_{\mu_L}}-m_{\nu_{e_L}}+R_L}{2m_{\nu_{e_L}\nu_{\mu_L}}},\quad
 {\rm with}\quad R_L^2=(m_{\nu_{e_L}}-m_{\nu_{\mu_L}})^2+4m_{\nu_{e_L}\nu_{\mu_L}}^2,
\end{equation}
and thus the neutrino fields with definite masses $\nu_1$ and
$\nu_2$ have respectively the following masses
\begin{align}
 m_{1}&=\frac{1}{2}(m_{\nu_{e_L}}+m_{\nu_{\mu_L}}-R_L),\\
 m_{2}&=\frac{1}{2}(m_{\nu_{e_L}}+m_{\nu_{\mu_L}}+R_L)e^{-i\alpha_L}.
\end{align}

We observe that in the expression for $m_2$ appears the factor
$\exp\{-i\alpha_L\}$ which suggest that this mass could be
complex. However the diagonalization given by (\ref{eq:DONM}) is
not completely right because $M_L$ is a symmetric matrix. So from
(\ref{eq:DONM}) the diagonalization should be of the form
\begin{align}
&M_{D_L}^2=U_L^\dag M_L^\dag M_L U_L,
\end{align}
where we have considered that this matrix is hermitic, i. e.
$M_L^2\equiv M_L^\dag M_L$. So the values $m_1$ and $m_2$ are the
quadratic roots of the eigenvalues of $M_L^2$. This last result
implies that these eigenvalues can be multiplied by a complex
phase.

The expression (\ref{eq:CMN}) gives the mixing of the flavor
neutrino fields in terms of the neutrino fields with definite
masses. The neutrino fields with definite masses $\nu_1$ and
$\nu_2$ obey Majorana field equations of the form
\begin{subequations}
\begin{align}
i\bar\sigma^{\mu}\partial_{\mu}\nu_{1}&=-im_{1}\sigma_{2}\nu_{1}^{*},\\
i\bar\sigma^{\mu}\partial_{\mu}\nu_{2}&=-im_{2}e^{-i\alpha}\sigma_{2}\nu_{2}^{*}.
\end{align}
\end{subequations}
With the purpose of eliminating the phase $\alpha_L$ from the last
equation of motion, we can make the following phase transformation
$\nu_2\rightarrow \exp\{-\frac{i\alpha_L}{2}\}\nu_2$. Now the
unitary matrix can be written as
\begin{align}\label{eq:MUD}
 U_L=\frac{1}{\sqrt{1+\Lambda_L^2}}\begin{pmatrix}
                                \Lambda_L & e^{-i\alpha_L}\\
                -1 & \Lambda_L e^{-i\alpha_L}
                               \end{pmatrix}.
\end{align}
We observe that the phase $\exp\{-i\alpha_L\}$ was eliminated from
the last equation of motion but not from the unitarian matrix
$U_l$. So it proves that this phase is physical and should be
involved in some processes. This phase could play an important
role in the case of doublet beta decay process.

Following a similar procedure for the right-handed Majorana
neutrinos, we find that the definite-mass neutrino field doublet
$n_2$ given by (\ref{eq:NL4}) is related to the flavor
right-handed neutrino doublet $\nu^c_R$ given by (\ref{eq:NL2}) by
mean of
\begin{equation}\label{eq:CMNR}
\nu^c_R=U_R n_2,
\end{equation}
where the unitary matrix $U_R$ is given by
\begin{align}\label{eq:MUDR}
 U_R=\frac{1}{\sqrt{1+\Lambda_R^2}}\begin{pmatrix}
                                \Lambda_R & e^{-i\alpha_R}\\
                -1 & \Lambda_R e^{-i\alpha_R}
                               \end{pmatrix}.
\end{align}
where $\Lambda_R$ is given by
\begin{equation} \label{eq:LambdaR}
 \Lambda_R =\frac{m_{\nu^c_{\mu_R}}-m_{\nu^c_{e_R}}+R_R}{2m_{\nu^c_{e_R}\nu^c_{\mu_R}}},\quad
 {\rm with}\quad
 R_R^2=(m_{\nu^c_{e_R}}-m_{\nu^c_{\mu_R}})^2+4m_{\nu^c_{e_R}\nu^c_{\mu_R}}^2.
\end{equation}

 Next we consider the canonical quantization of the
neutrino fields with definite mass by stating the anticonmutation
relations given by
$\llav{\des{\nu}{a}(\vec{r},h),\crea{\nu}{b}(\vec{r}',h')}=
\delta_{ab}\delta^3(\vec{r}-\vec{r}')$,
$\llav{\des{\nu}{a}(\vec{r},h),\des{\nu}{b}(\vec{r}',h')}= 0$ and
$\llav{\crea{\nu}{a}(\vec{r},h),\crea{\nu}{b}(\vec{r}',h')}= 0$,
where $a,b=1,2,3,4$ represent neutrino mass states. Each one of
the definite-mass neutrino field operators $\hat\nu_a(x)$ obeys a
Majorana equation. It is possible to expand each one of these
field operators on a plane-wave basis set as was shown in
(\ref{eq:EONM})
\begin{align}\label{eq:FEOFDM}
\hat\nu_a(x)=\int
\frac{d^3p}{(2\pi)^{3/2}(2E_a)^{1/2}}&\sum_{h=\pm1}\left[\sqrt{E_a-h
\abs{\vec{p}}}\,\des{a}{a}(\vec{p},h)\chi^{h}(\vec{p})e^{-ip\cdot
x}\right.\notag\\
&\qquad\quad\left.-h\sqrt{E_a+h\abs{\vec{p}}}\,\crea{a}{a}(\vec{p},h)
\chi^{-h}(\vec{p})e^{ip\cdot
x}\right],
\end{align}
where $E_a^2=\abs{\vec{p}}^2+m_a^2$ is the energy of the neutrino
field with definite mass which is tagged by $a=1,2,3,4$.

The flavor neutrino field operators tagged by $\hat \nu_\alpha$
are defined as superposition of the definite-mass neutrino field
operators $\hat \nu_a$ given by (\ref{eq:FEOFDM}) through the
expression
\begin{align}
 \des{\nu}{\alpha}(x)=\sum_aU_{\alpha a}\des{\nu}{a}(x),
\end{align}
where $U$ is the unitarian matrix defined by (\ref{eq:MUD}) for
left-handed neutrinos and by (\ref{eq:MUDR}) for right-handed
neutrinos, meanwhile neutrino flavor states $\ket{\nu_\alpha}$ are
defined in terms of the neutrino mass states $\ket{\nu_a}$ as
\begin{align}\label{eq:DefEstSab}
 \ket{\nu_\alpha}=\sum_aU^*_{\alpha a}\ket{\nu_a}.
\end{align}
Thus we have found a relation between neutrino flavor states and
neutrino mass states using operator fields. As flavor states are
physical states since they could be detected in interaction
processes, flavor states are non-stationary. So their temporal
evolution gives the probability of transition between them.
Therefore, this probability describes Majorana neutrino
oscillations studied as follows.


\section{Left-handed neutrino oscillations}

Now we will focuss our interest in the description of left-handed
neutrino oscillations in vacuum from a cinematical point of view.
For this reason we will not consider in detail the weak
interaction processes involved in the creation and detection of
left-handed neutrinos. However, these processes are manifested
when boundary conditions are imposed in the probability amplitude
of transition between two neutrino flavor states. We suppose that
a neutrino with a specific flavor is created in a point of
space-time $x_0^\mu=(t_0,\vec{r}_0)$ as a result of a certain weak
interaction process. We will determine the probability amplitude
to find out the neutrino with another flavor in a different point
of space-time $x^\mu=(t,\vec{r})$. We assume that neutrinos are
created under the same production process with different values of
energy and momentum. These dynamical quantities are related among
themselves under the specific production process.\\

The initial left-handed neutrino flavor state in the production
time ($t_0$) corresponds to the following superposition of
neutrino mass states
\begin{align}
\ket{\nu_L(t_0)}=A\ket{\nu_1}+B\ket{\nu_2}.
\end{align}
where $\abs{A}^2+\abs{B}^2=1$. Each of these neutrino mass states
has associated a specific four-momentum. We assume that in the
production point it was created a left-handed electronic neutrino
with each massive field having a four-momentum given by
$p_a^\mu=(E_a,\vec{p}_a)$, with $a=1,2$. The initial left-handed
electronic neutrino state satisfying the condition
$\abs{A}^2+\abs{B}^2=1$ is written as
\begin{align}\label{eq:EI}
\ket{\nu_L(t_0)}=\sum_{h=\pm 1}\frac{\Lambda_L
}{\sqrt{1+\Lambda_L^2}}\ket{\nu_1}+\frac{e^{-i\alpha_L}}{\sqrt{1+
\Lambda_L^2}}\ket{\nu_2},
\end{align}
where the sum over helicities is taken over the neutrino mass
states. This sum over helicities must be considered to describe
appropriately the initial left-handed neutrino flavor state
because the helicity is a property which is not directly measured
in the experiments. The manner as the electronic left-handed
neutrino state has been built in the production point is in
agreement with the experimental fact that left-handed neutrinos
are ultra-relativistic.

The neutrino mass states involve in the superposition given by
(\ref{eq:EI}) are obtained from the vacuum state as
$\ket{\nu_a}=e^{ip_ax_0}\crea{a}{a}(\vec{p}_a,h)\ket{0}$, where we
have included the phase factor $\exp\{ip_ax_0\}$. This phase
factor gives us information about the four-space time where the
left-handed neutrino was created. The probability amplitudes for
transitions to electronic and muonic left-handed neutrinos are
respectively given by
\begin{align} \label{eq:nuapt}
\nu_{e_L}(X)&\equiv\bra{0}\des{\nu}{e_L}(x)\ket{\nu_L(t_0)}\notag\\
&=\frac{1}{(2\pi)^{3/2}}\sum_{h}\left\{
\frac{\Lambda_L^2}{1+\Lambda_L^2}\sqrt{\frac{E_1-h\abs{\vec
p}_1}{2E_1}} e^{-ip_1X}\chi^h(\vec{p}_1)
+\frac{1}{1+\Lambda_L^2}\sqrt{\frac{E_2-h \abs{\vec
p}_2}{2E_2}}e^{-ip_2X}\chi^h(\vec{p}_2)\right\},
\end{align}
\begin{align} \label{eq:eapt}
\nu_{\mu_L}(X)&\equiv\bra{0}\hat\nu_{\mu_L}(x)\ket{\nu_L(t_0)}\notag\\
           &=\frac{1}{(2\pi)^{3/2}}\sum_{h}\left\{
-\frac{\Lambda_L}{1+\Lambda_L^2}\sqrt{\frac{E_1-h\abs{\vec
p}_1}{2E_1}} e^{-ip_1X}\chi^h(\vec{p}_1)
+\frac{\Lambda_L}{1+\Lambda_L^2} \sqrt{\frac{E_2-h\abs{\vec
p}_2}{2E_2}}e^{-ip_2X}\chi^h(\vec{p}_2)\right\},
\end{align}
where we have used some expansions over the Majorana fields and we
have taken  $X\equiv x-x_0$ which corresponds to a four-vector
associated to the distance and time of neutrino propagation. The
probability densities $\rho_{\nu_\alpha
L}(X)=\abs{\nu_{\alpha_L}(X)}^2$ respectively are
\begin{align} \label{eq:nudpt}
  \rho_{\nu_{eL}}(X)&=\frac{1}{(2\pi)^{3}}\frac{1}{(1+\Lambda_L^2)^2}\times\notag\\
                 &\quad\left\{1+\Lambda_L^4+\frac{\Lambda_L^2}{(E_1E_2)^{1/2}}
                 \sum_h\sqrt{(E_1-h\abs{\vec p}_1)(E_2-h\abs{\vec p}_2)}\right.
                 \notag\\&\quad \left.-\frac{(2\Lambda_L)^2}{2(E_1E_2)^{1/2}}
                 \sum_h\sqrt{(E_1-h\abs{\vec p}_1)(E_2-h\abs{\vec p}_2)}
                 \sin^2\corc{\frac{p_1-p_2}{2}X}\right\},
\end{align}
\begin{align}\label{eq:edpt}
 \rho_{\nu_{\mu L}}(X)&=\frac{1}{(2\pi)^{3}}\frac{1}{(1+\Lambda_L^2)^2}\times\notag\\
                   &\quad\left\{2\Lambda_L^2-\frac{\Lambda_L^2}{(E_1E_2)^{1/2}}\sum_h
                   \sqrt{(E_1-h\abs{\vec p}_1)(E_2-h\abs{\vec p}_2)}\right.
                   \notag\\&\quad
\left.+\frac{(2\Lambda_L)^2}{2(E_1E_2)^{1/2}}\sum_h\sqrt{(E_1-h\abs{\vec
p}_1)(E_2-h\abs{\vec
p}_2)}\sin^2\corc{\frac{p_1-p_2}{2}X}\right\},
\end{align}
where we have taken into account that spinors $\chi^h(\vec{p}_1)$
and $\chi^h(\vec{p}_2)$ are the same because vectors $\vec{p}_1$ and
$\vec{p}_2$ are co-linear. The probability densities
(\ref{eq:nudpt}) and (\ref{eq:edpt}) that we have found present a
serious problem. If we fix $X=0$ into (\ref{eq:nudpt}) and
(\ref{eq:edpt}) we find that
\begin{align} \label{eq:nudptxo}
\rho_{\nu_{eL}}(X=0)=\frac{1}{(2\pi)^{3}}\frac{1}{(1+\Lambda_L^2)^2}
\left\{1+\Lambda_L^4+\frac{\Lambda_L^2}{(E_1E_2)^{1/2}}\sum_h
\sqrt{(E_1-h\abs{\vec p}_1)(E_2-h\abs{\vec p}_2)}\right\},
\end{align}
\begin{align}\label{eq:edptxo}
 \rho_{\nu_{\mu L}}(X=0)=\frac{1}{(2\pi)^{3}}\frac{1}{(1+\Lambda_L^2)^2}
 \left\{2\Lambda_L^2-\frac{\Lambda_L^2}{(E_1E_2)^{1/2}}\sum_h
 \sqrt{(E_1-h\abs{\vec p}_1)(E_2-h\abs{\vec p}_2)}\right\},
\end{align}
and we observe that the probability density (\ref{eq:edptxo}) can
be different from zero, i. e. it can exist a muonic neutrino in
the production point which disagrees with the initial conditions.
The origin of this problem is related to the weak state definition
\eqref{eq:DefEstSab} that we have used before. As it was
previously mentioned into the introduction, the flavor definition
\eqref{eq:DefEstSab} is not complectly consistent and it is
necessary to define appropriate flavor states \cite{Giunti2}.


\subsection{Ultra-relativistic limit: Left-handed neutrino oscillations}

This problem can be solved by taking an approximation in the
probability densities (\ref{eq:nudpt}) and (\ref{eq:edpt}) based
on the fact that left-handed neutrinos are ultra-relativistic
particles because their masses are very small. Here we consider
energy and momentum different for every mass state. In general we
can write
\begin{align}\label{eq:CEM1}
  \abs{\vec p}_a^2&=E^2-\xi m_a^2+\zeta m_a^4,
\end{align}
\begin{align}\label{eq:CEM2}
  E_a^2&=E^2+(1-\xi) m_a^2+\zeta m_a^4,
\end{align}
where the parameters $\xi$ and $\zeta$ are determined in the
production process and $E$ is the energy for the case in which
neutrinos were massless. For instance, for the pion decay process we
have
\begin{align}
  E&=\frac{m_\pi}{2}\corc{1-\frac{{m'}_{\mu}^2}{m_\pi^2}},\\
  \xi&=\frac{1}{2}\corc{1+\frac{{m'}_{\mu}^2}{m_\pi^2}},\qquad
  \zeta=\frac{1}{4m_\pi^2},
\end{align}
where ${m'}_\mu$ is the muon mass and $m_\pi$ is the pion mass.
Because for the ultra-relativistic limit $m_a\rightarrow 0$, we can
approximate the expressions (\ref{eq:CEM1}) and (\ref{eq:CEM2}) to
\begin{align}\label{eq:AUR}
  \abs{\vec p}_a&\approx E-\xi\frac{m_a^2}{2E},\\
  E_a&\approx E+(1-\xi)\frac{m_a^2}{2E}.
\end{align}
Now it is possible to prove that the right side of the relation
\begin{align}\label{eq:AFS}
 \frac{1}{2(E_1E_2)^{1/2}}\sum_{h=\pm1}\sqrt{(E_1-h\abs{\vec p}_1)
 (E_2-h\abs{\vec p}_2)}\approx 1-\frac{(m_1-m_2)^2}{8E^2},
\end{align}
can be approximated to the unit because $(\delta
m_{12}/E)^2\approx 0$, where $\delta m_{12}= m_1-m_2$. On the
other hand, neutrino propagation time $T$ is not measured in
neutrino experiments \cite{GK07,Beuthe,Giunti1}. In this kind of
experiments is measured the distance $L$ between the neutrino
source and the detector. By this reason, it can be possible to
find a analytical expression that establishes a relation between
$T$ and the propagation distance $L=\abs{\vec{L}}$. In our
approach using plane waves, for the ultra-relativistic limit we
can write
\begin{align}\label{eq:RTL}
L\approx T.
\end{align}
This relation implies that the propagation distance and the
propagation time for neutrinos are approximately equal because in
the ultra-relativistic limit a neutrino mass state has a mass too
small and its velocity of propagation $v_k$ is approximately equal
to speed velocity $c=1$, i. e. $v_k \approx 1$. However, a most
precise relation between $L$ and $T$ must be described by an
expression that should include explicitly the velocities of the
two neutrino mass states involved in such a way that this
expression for the ultra-relativistic limit should lead to
\eqref{eq:RTL}.\\

So for the ultra-relativistic limit the probability densities
(\ref{eq:nudpt}) and (\ref{eq:edpt}) can be written as
\begin{align}\label{eq:nudpt2}
  \rho_{\nu_{e L}}(L)&=\frac{1}{(2\pi)^{3}}\left\{1-\corc{\frac{2\Lambda_L}
  {1+\Lambda_L^2}}^2\sin^2\corc{\frac{\Delta m_{12}^2}{4E}L}\right\},
\end{align}
\begin{align}\label{eq:edpt2}
  \rho_{\nu_{\mu L}}(L)&=\frac{1}{(2\pi)^{3}}\corc{\frac{2\Lambda_L}
  {1+\Lambda_L^2}}^2\sin^2\corc{\frac{\Delta m_{12}^2}{4E}L},
\end{align}
where we have used $L\approx T$ and $\Delta m_{12}^2\equiv
m_1^2-m_2^2$. Under this approximation it is clear that these
probability density does not depend from the production process
due to that there is no dependence from $\xi$. Thus these
probability densities satisfy the boundary conditions that we have
imposed.

In the next we will prove that the probability densities
(\ref{eq:nudpt2}) and (\ref{eq:edpt2}) have the form of the
standard probability densities for neutrino oscillations. In the
context of the standard formalism of neutrino oscillations
(assuming CP conservation), for the two generation case
considering here, the representation of the unitary matrix $U_L$
that appears into the expression (\ref{eq:CMN}) is given by
\cite{CK93}
\begin{align}\label{eq:MUDSR}
 U_L=\begin{pmatrix}
 \cos\theta_L & \sin\theta_L\\
-\sin\theta_L &  \cos\theta_L
 \end{pmatrix},
\end{align}
where $\theta_L$ is the mixing angle. If we compare the unitary
matrix given by (\ref{eq:MUD}) with the one given by
(\ref{eq:MUDSR}), we observe that $\cos\theta_L = \Lambda_L
/(\sqrt{1+\Lambda_L^2})$ and then it is very easy to obtain that
\begin{equation}\label{eq:thlamb}
\sin 2\theta_L=\frac{2\Lambda_L}{1+\Lambda_L^2}.
\end{equation}
Substituting (\ref{eq:thlamb}) into (\ref{eq:nudpt2}) and
(\ref{eq:edpt2}), we obtain the expressions
\begin{align}\label{eq:nudpt3}
  \rho_{\nu_{e L}}(L)&=\frac{1}{(2\pi)^{3}}\left\{1-\sin^2(2\theta_L)
  \sin^2\corc{\frac{\Delta m_{12}^2}{4E}L}\right\},
\end{align}
\begin{align}\label{eq:edpt3}
  \rho_{\nu_{\mu L}}(L)&=\frac{1}{(2\pi)^{3}}\sin^2(2\theta_L)
  \sin^2\corc{\frac{\Delta m_{12}^2}{4E}L},
\end{align}
which are the standard probability densities for left-handed
neutrino oscillations in the two flavor case \cite{CK93}.


\section{Right-handed neutrino oscillations}

The initial right-handed neutrino flavor state in the production
time ($t_0$) corresponds to the following superposition of
neutrino mass states
\begin{align}
\ket{\nu^c(t_0)}=C\ket{\nu_3}+D\ket{\nu_4}.
\end{align}
where $\abs{C}^2+\abs{D}^2=1$. Each of these neutrino mass states
has associated a specific four-momentum. We assume that in the
production point it was created a right-handed electronic neutrino
with each massive field having a four-momentum given by
$p_a^\mu=(E_a,\vec{p}_a)$, with $a=3,4$. The initial right-handed
electronic neutrino state satisfying the condition
$\abs{C}^2+\abs{D}^2=1$ is written as
\begin{align}\label{eq:EIR}
\ket{\nu^c_R(t_0)}=\sum_{h=\pm 1}\frac{\Lambda_R
}{\sqrt{1+\Lambda_R^2}}\ket{\nu_3}+\frac{e^{-i\alpha_R}}{\sqrt{1+
\Lambda_R^2}}\ket{\nu_4},
\end{align}
where the sum over helicities is taken over the neutrino mass
states.

The probability amplitudes for transitions to electronic and
muonic right-handed neutrinos are respectively given by
\begin{align} \label{eq:nuaptR}
\nu^c_{e_R}(X)&\equiv\bra{0}\des{\nu^c}{e_R}(x)\ket{\nu^c_R(t_0)}\notag\\
&=\frac{1}{(2\pi)^{3/2}}\sum_{h}\left\{
\frac{\Lambda_R^2}{1+\Lambda_R^2}\sqrt{\frac{E_3-h\abs{\vec
p}_3}{2E_3}} e^{-ip_3X}\chi^h(\vec{p}_3)
+\frac{1}{1+\Lambda_R^2}\sqrt{\frac{E_4-h \abs{\vec
p}_4}{2E_4}}e^{-ip_4X}\chi^h(\vec{p}_4)\right\},
\end{align}
\begin{align} \label{eq:eaptR}
\nu^c_{\mu_R}(X)&\equiv\bra{0}\hat\nu^c_{\mu_R}(x)\ket{\nu^c_R(t_0)}\notag\\
           &=\frac{1}{(2\pi)^{3/2}}\sum_{h}\left\{
-\frac{\Lambda_R}{1+\Lambda_R^2}\sqrt{\frac{E_3-h\abs{\vec
p}_3}{2E_3}} e^{-ip_3X}\chi^h(\vec{p}_3)
+\frac{\Lambda_R}{1+\Lambda_R^2} \sqrt{\frac{E_4-h\abs{\vec
p}_4}{2E_4}}e^{-ip_4X}\chi^h(\vec{p}_4)\right\}.
\end{align}
The probability densities $\rho_{\nu_\alpha
R}(X)=\abs{\nu^c_{\alpha_R}(X)}^2$ respectively are
\begin{align} \label{eq:nudptR}
  \rho_{\nu^c_{eR}}(X)&=\frac{1}{(2\pi)^{3}}\frac{1}{(1+\Lambda_R^2)^2}\times\notag\\
                 &\quad\left\{1+\Lambda_R^4+\frac{\Lambda_R^2}{(E_3E_4)^{1/2}}
                 \sum_h\sqrt{(E_3-h\abs{\vec p}_3)(E_4-h\abs{\vec p}_4)}\right.
                 \notag\\&\quad \left.-\frac{(2\Lambda_R)^2}{2(E_3E_4)^{1/2}}
                 \sum_h\sqrt{(E_3-h\abs{\vec p}_3)(E_4-h\abs{\vec p}_4)}
                 \sin^2\corc{\frac{p_3-p_4}{2}X}\right\},
\end{align}
\begin{align}\label{eq:edptR}
 \rho_{\nu^c_{\mu R}}(X)&=\frac{1}{(2\pi)^{3}}\frac{1}{(1+\Lambda_R^2)^2}\times\notag\\
                   &\quad\left\{2\Lambda_R^2-\frac{\Lambda_R^2}{(E_3E_4)^{1/2}}\sum_h
                   \sqrt{(E_3-h\abs{\vec p}_3)(E_4-h\abs{\vec p}_4)}\right.
                   \notag\\&\quad
\left.+\frac{(2\Lambda_R)^2}{2(E_3E_4)^{1/2}}\sum_h\sqrt{(E_3-h\abs{\vec
p}_3)(E_4-h\abs{\vec
p}_4)}\sin^2\corc{\frac{p_3-p_4}{2}X}\right\},
\end{align}
where we have taken into account that spinors $\chi^h(\vec{p}_3)$
and $\chi^h(\vec{p}_4)$ are the same because vectors $\vec{p}_3$
and $\vec{p}_4$ are co-linear. The probability densities
(\ref{eq:nudptR}) and (\ref{eq:edptR}) that we have found present
a serious problem. If we fix $X=0$ into (\ref{eq:nudptR}) and
(\ref{eq:edptR}) we find that
\begin{align} \label{eq:nudptxoR}
\rho_{\nu^c_{e
R}}(X=0)=\frac{1}{(2\pi)^{3}}\frac{1}{(1+\Lambda_R^2)^2}
\left\{1+\Lambda_R^4+\frac{\Lambda_R^2}{(E_3E_4)^{1/2}}\sum_h
\sqrt{(E_3-h\abs{\vec p}_3)(E_4-h\abs{\vec p}_4)}\right\},
\end{align}
\begin{align}\label{eq:edptxoR}
 \rho_{\nu_{\mu R}}(X=0)=\frac{1}{(2\pi)^{3}}\frac{1}{(1+\Lambda_R^2)^2}
 \left\{2\Lambda_R^2-\frac{\Lambda_R^2}{(E_3E_4)^{1/2}}\sum_h
 \sqrt{(E_3-h\abs{\vec p}_3)(E_4-h\abs{\vec p}_4)}\right\},
\end{align}
and we observe that the probability density (\ref{eq:edptxoR}) can
be different from zero.


\subsection{Non-relativistic limit: Right-handed neutrino oscillations}

This problem can be solved by taking an approximation in the
probability densities (\ref{eq:nudptR}) and (\ref{eq:edptR}) based
on the fact that right-handed neutrinos are non-relativistic
particles because their masses are very large. By this reason, we
take the non-relativistica approximation, i. e. $m_a\gg p_a$. So
we have
\begin{align}
E_a\approx m_a+\frac{\abs{\vec p}_a^2}{2m_a}.
\end{align}
Therefore, we suppose that heavy right-handed Majorana neutrinos
obey simply the relativistic dispersion relation. So we obtain the
following approximation
\begin{align}
\frac{1}{2(E_3E_4)^{1/2}}\sum_{h=\pm1}\sqrt{(E_3-h\abs{\vec
p}_3)(E_4-h\abs{\vec p}_4)}=
1+\frac{v_3v_4}{2}-\frac{1}{8}(v_3^2+v_4^2)(v_3+v_4)^2+\cdots,
\end{align}
where the non-relativistic velocity of the neutrino is $v_i\equiv
\frac{\abs{\vec p}_i}{m_i}\approx 0$, meanwhile the phase is
approximated to
\begin{align}
(E_3-E_4)T-(p_3-p_4)L\approx\Delta m_{34}T,
\end{align}
with $\Delta m_{34}\equiv m_3-m_4$. So the probability densities
of transition are given by
\begin{align}
  \rho_{\nu^c_{e R}}(T)&=\frac{1}{(2\pi)^{3}}\left\{1-\corc{\frac{2\Lambda_R}
  {1+\Lambda_R^2}}^2\sin^2\corc{\frac{\Delta m_{34}}{2}T}\right\},
  \label{eq:rhMnpde}\\
  \rho_{\nu^c_{\mu R}}(T)&=\frac{1}{(2\pi)^{3}}\corc{\frac{2\Lambda_R}{1+
  \Lambda_R^2}}^2\sin^2\corc{\frac{\Delta m_{34}}{2}T}\label{eq:rhMnpdmu},
\end{align}
where $\Lambda_R$ is given by (\ref{eq:LambdaR}). The last
probability densities satisfy the normalization and boundary
conditions. Unlikely to the case of left-handed neutrino
oscillations described by (\ref{eq:nudpt}) and (\ref{eq:edpt}),
the argument of the periodic function for the right-handed
neutrino oscillations depends on the linear mass difference
$\Delta m_{34}$ and the propagation time $T$. The description of
heavy right-handed neutrino oscillations that we present here
could be of interest in cosmological problems \cite{Gershtein}. As
it has been proposed in the literature, heavy-heavy neutrino
oscillations could be responsible for the baryon asymmetry of the
universe through a leptogenesis mechanism
\cite{AkhRubaSmir,Volkas}. But it should be noted that if the
propagation of heavy right-handed neutrinos is considered as
superpositions of mass-eigenstate wave packets \cite{Kim}, then
the oscillations do not take place because the coherence is not
preserved: in other words, the oscillation length is comparable or
larger than the coherence length of the right-handed neutrino
system \cite{Kim}.


\section{Conclusions}

In this work we have studied neutrino oscillations in vacuum
between two flavor states considering neutrinos as Majorana
fermions. We have performed this study for the case of flavor
states constructed as superpositions of mass states extending the
Sassaroli model which describes Majorana neutrino oscillations by
considering neutrino mass states as plane waves with specific
momenta. In the context of a type I seesaw scenario which leads to
get light left-handed and heavy right-handed Majorana neutrinos,
the main contribution of this work has been to obtain in a same
formalism the probability densities which describe the
oscillations for light left-handed neutrinos (ultrarelativistic
limit) and for heavy right-handed neutrinos (non-relativistic
limit). In this work we have performed the canonical quantization
procedure for Majorana neutrino fields of definite masses and then
we have written the neutrino flavor states as superpositions of
mass states using quantum field operators. We have calculated the
probability amplitude of transition between two different neutrino
flavor states for the light and heavy neutrino cases and we have
established normalization and boundary conditions for the
probability density. After the ultra-relativistic limit was taken
in the probability densities for the left-handed neutrino case
lead to the standard probability densities which describe light
neutrino oscillations. For the right-handed neutrino case, the
expressions describing heavy neutrino oscillations in the
non-relativistic limit were different respect to the ones of the
standard neutrino oscillations. However, the right-handed neutrino
oscillations are phenomenologically restricted as is shown when
the propagation of heavy neutrinos is considered as superpositions
of mass-eigenstate wave packets \cite{Kim}.

This work establish a framework to study Majorana neutrino
oscillations for the case where mass states are described by
Gaussian wave packets as will be presented in a forthcoming work
\cite{PQ1}. The wave packet treatment is necessary owing to the
neutrinos are produced in weak interaction processes without a
specific momenta. Additionally the plane wave treatment can not
describe production and detection localized processes as occur in
neutrino oscillations.

\section*{Acknowledgments}

C. J. Quimbay thanks DIB by the financial support received through
the research project "Propiedades electromagn\'eticas y de
oscilaci\'on de neutrinos de Majorana y de Dirac". C. J. Quimbay
thanks also Vicerrector\'ia de Investigaciones of Universidad
Nacional de Colombia by the financial support received through the
research grant "Teor\'ia de Campos Cu\'anticos aplicada a sistemas
de la F\'isica de Part\'iculas, de la F\'isica de la Materia
Condensada y a la descripci\'on de propiedades del grafeno".


\end{document}